\documentclass[twocolumn,showpacs,prl,aps,10pt,balancelastpage]{revtex4-1}
\usepackage{calc}
\usepackage{graphicx}
\usepackage[squaren]{SIunits}
\usepackage{amsmath}
\usepackage{color}
\usepackage{tikz}
\usepackage{lmodern}
\usepackage{amssymb}
\usepackage{comment}
\usepackage{enumitem} % no spaces in list
\usepackage{soul}
\usepackage[colorlinks=true, linkcolor=blue, citecolor=blue, filecolor=blue, urlcolor=blue, bookmarks=true, bookmarksopen=true, bookmarksopenlevel=3, plainpages=false, pdfpagelabels=true]{hyperref}
\definecolor{darkgreen}{rgb}{0,0.4,0}
\definecolor{pink}{rgb}{0.7,0.0,0.6}
\definecolor{dkred}{rgb}{0.7,0.0,0.0}
\newcommand{\www}{0.999\columnwidth}

\newcommand{\minus}{\scalebox{0.79}{$-$}}

\begin{document}

%\title {Towards the Understanding of Ultrafast Magnetization Dynamics}
%\title{Driving ferromagnets into a critical phase}
\title{Driving ferromagnets into a critical region of a magnetic phase diagram}
\author{B. Y. Mueller}
\email {bmueller@physik.uni-kl.de}
\author{B. Rethfeld}
\affiliation{Department of Physics and Research Center OPTIMAS, University of Kaiserslautern, Erwin-Schr\"odinger-Str.~46, 67653 Kaiserslautern, Germany}

\date{\today}
\pacs{75.78.-n, 75.78.Jp, 05.70.Ln}

%----------------------------------ABSTRACT-------------------------------------------------------------------------
\begin{abstract}
Exciting a ferromagnetic sample with 
an ultrashort laser pulse leads to a quenching of the magnetization on a
subpicosecond timescale. 
%Utilizing findings of previous 
%studies applying a detailed 
%microscopic Boltzmann approach
On the basis of the equilibration of 
intensive thermodynamic variables
%of the majority and minority electrons
we establish a powerful model
to describe the demagnetization dynamics.
%This model %leads to a detailed understanding
We demonstrate that the magnetization dynamics is
mainly driven by the equilibration of chemical potentials.
%of the initial demagnetization and 
The minimum of magnetization is revealed as a transient 
electronic equilibrium state. 
Our method identifies the slowing down of
ultrafast magnetization dynamics by 
%the crossing of 
a critical region within a magnetic phase
diagram.

%
%Several models compete to explain the underlying microscopic mechanisms,
%in particular spin-flip scattering and superdiffusive transport have 
%been invoked as hot candidates. 
%%Since this discovery, several models for the microscopic mechanism have %%been proposed, 
%%for instance, spin-flip scattering and superdiffusive transport.
%Today it seems that both processes
%contribute to the demagnetization, depending on the sample properties. 
%%
%Applying a spin-resolved Boltzmann equation, 
%we find that spin-flip scattering 
%is driven by the equilibration of temperatures 
%and chemical potentials of spin-up and 
%spin-down electrons.
%We further show that the dynamical exchange splitting
%yields a much larger quenching of magnetization than 
%in the case of a fixed band structure.
%%
%With these basic assumptions, we set up a system of equations as a 
%simplified phenomenological model to describe the demagnetization 
%governed by these temperatures and chemical potential equilibrations. 
%%This opens the possibility to describe transport effects in parallel %with 
%%spin-flip events and allows for analytical predictions 
%%of the demagnetization. 
\end{abstract}

\maketitle

The strong increase of computational power within the last thirty years
has also boosted the need for large and fast data storage.
However,  the physical speed limits of conventional magnetic recording, 
which are on the order of nanoseconds, are nowadays reached \cite{McDaniel2005}.
A promising enhancement lies in a subpicosecond
change of magnetization, as has been found 
in 1996 by exciting a ferromagnetic material with an ultrashort laser pulse \cite{Beaurepaire1996}.
Though, a detailed understanding of the underlying physical processes
of this ultrafast demagnetization is still lacking
and several models compete, hampering the further development~\cite{Schellekens2013a,Schellekens2013b,Illg2013,Carva2011}. 

The most promising concepts are based on 
superdiffusive spin transport~\cite{Battiato2010,Eschenlohr2013,Turgut2013,Kaltenborn2012}
or
Elliott-Yafet (EY) spin-flip processes~\cite{Essert2011,Koopmans2010,MuellerBY,MuellerPRL,Krauss2009,Steil2010,Koopmans2005,Fahnle2010,Walowski2008,Illg2013,Carva2011,Carpene2008,Koopmans2005,Roth2012}.
%Both effects play a significant role and their influence depends on the the sample geometry~%\cite{Turgut2013, Schellekens2013a}.
It has been shown experimentally, that
both processes contribute to the 
magnetization dynamics, depending on the sample geometry~\cite{Turgut2013, Schellekens2013a}.
On the one hand, 
superdiffusive transport dominates on bulk and multilayer systems
and has been successfully compared to experiments~\cite{Turgut2013,Eschenlohr2013}.
On the other hand, 
EY spin-flip scattering has been investigated with kinetic models
and reproduces the magnetization dynamics for thin films \cite{MuellerPRL,Krauss2009,Steil2010}.
Due to the complexity of the methods, 
temperature-based models have been proposed, 
like the microscopic three temperature model (M3TM) ~\cite{Koopmans2010,Roth2012}.
Recently, it has been shown, 
that this simplification is justified, 
despite of an ultrafast laser excitation \cite{MuellerPRL}.
%\pink{superdiffusive immer noch wichtiger? }

In this Letter, we derive a $\mu$T model ($\mu$TM), 
which traces the dynamics and the equilibration of 
temperatures and chemical potentials 
of the electron subsystems
simultaneously.
The essential concepts of the $\mu$TM  are based on a kinetic approach~\cite{MuellerPRL,MuellerBY,Krauss2009},
including EY-type spin-flip scattering and a dynamic exchange splitting~\cite{MuellerPRL,Essert2011}.
The $\mu$TM reproduces the experimental 
magnetization curves for different laser fluences.
We find that the equilibration of chemical potentials
drives 
 the 
dynamics of the magnetization 
 and 
the magnetization minimum is revealed as
a transient equilibrium state within a magnetic phase diagram.
We
identify a critical region within this phase diagram:
%\pink{Ich bestehe nicht auf ,,main message'', wir brauchen aber eine 
%Ueberleitung.} Nein. Wir machen den Punkt 1 und dann Punkt 2, zack und zack.
For certain fluences, the material is driven into 
this region, causing an extreme deceleration 
of the magnetization dynamics.
This finding confirms the experimental observation of 
a critical slowing down~\cite{Koopmans2010,Roth2012}.
Unlike the M3TM, we trace the dynamics of minority and majority electron densities explicitly,
which opens the possibility to extend the model
 for superdiffusive transport effects.

A general matrix formulation of a time- and space-dependent 
coupled transport equation is given by 
\begin{align}
	C \frac{d}{d t}  \vec X=& 
	\nabla K \nabla \vec X + G\vec X  + \vec S
\label{XTM}\enspace,
 \end{align} 
where $\vec X$ is the vector of transient variables, 
$C$, $K$ and $G$ are matrices of capacities, transport and 
coupling, respectively, and $\vec S$ is the source vector. 
A representative of such equation system is the well-known 
two temperature model (TTM) \cite{Anisimov} 
of two coupled heat conduction equations.
In that case the vector of interest $\vec X$
consists of the respective electron and lattice temperature, 
$T_e$ and $T_{\ell}$,
the source vector contributes to the equation for
the electron energy, and the capacity matrix as well as the transport
matrix are diagonal matrices. The temperatures are 
coupled through an equilibration term, 
$\pm g \left(T_e-T_{\ell}\right)$, thus the coupling 
matrix $G$ contains also off-diagonal elements.
Here, $g$ is the electron--lattice coupling parameter.

%%%%%%%%%
%Jetzt Magnetismus!

In itinerant ferromagnets, 
%we apply an effective
%Stoner model where 
the electrons of majority and minority
spins can be treated separately.
%It has been shown that this description works well for nickel.
The temperatures of both electron types, denoted by 
$T_e^{\uparrow}$ and $T_e^{\downarrow}$, respectively, may differ.
Moreover, the
respective particle densities 
may change due to 
EY spin-flip processes and only their sum $n=n^{\uparrow} + n^{\downarrow}$ is constant.
Therefore, the chemical potentials $\mu^{\uparrow}$ and 
$\mu^{\downarrow}$ 
have to be considered as further 
variables of $\vec X$ in Eq.~\eqref{XTM}.
%, similar to the case of the density-dependent 
%two-temperature description
%of semiconductors\cite{vanDriel1987}. \red{evtl weglassen}
Further, in the frame of an effective Stoner model~\cite{NoltingMag, MuellerPRL}, 
the densities of states 
$D^{\uparrow}(E)$ and $D^{\downarrow}(E)$
of up-
and down electrons, respectively, 
are shifted by an exchange splitting $\Delta$. 
%This can be formally treated with a vanishing 
%$D^{\downarrow}(E, \Delta)$
%for low energies $E < \Delta$ of the minority electrons.
This exchange splitting is not constant but is directly coupled with the 
magnetization $m$ through the effective Coulomb interaction $U$
\cite{NoltingMag}.
In Ref.~\onlinecite{MuellerPRL} it was shown that 
the instantaneous feedback of the transient magnetization on the 
exchange splitting, 
\begin{align}
	\Delta(t)=U m(t)\enspace,
\label{exchange}
\end{align}
is essential for the quantitative description of 
demagnetization dynamics. 
The normalized 
magnetization $m(t)$ results from the transient particle density of 
each electron reservoir as
\begin{align}
\label{eq:m}
m(t)=\left(n^\uparrow(t)-n^\downarrow(t)\right)/n
\enspace.
\end{align}

%The particle density for a given Fermi distribution $f(E,T_e,\mu)$
%of temperature $T_e$ and 
%chemical potential $\mu$ in a material with density of 
%states $D(E,\Delta)$ is determined by the integration over the 
%distribution 
%\begin{align}
%\label{eq:N} 
%	n(T_e,\mu,\Delta)=\int D(E,\Delta)\, f(E,T_e,\mu)\, dE \enspace.
%\end{align}
%For the same distribution
%the internal energy density $u_e$ is determined
%by the second moment as 
%\begin{align}
%\label{eq:u} 
%	u_e(T_e,\mu,\Delta)=\int E \,D(E,\Delta)\, f(E,T_e,\mu)\, dE \enspace.
%\end{align}
The particle density 
$n^{\sigma}\left(T_e^{\sigma},\mu^{\sigma},m\right)$ 
and internal energy density
$u^{\sigma}_e\left(T_e^{\sigma},\mu^{\sigma},m\right)$
of the spin $\sigma\in\{\uparrow,\downarrow\}$
are calculated by the zeroth and first moment of 
the current Fermi distribution $f\left(E,T^{\sigma}_e,\mu^{\sigma}\right)$.
Thus, under the given conditions both, 
particle density and internal energy density, 
depend on the two intrinsic
variables $T^{\sigma}_e$ and $\mu^{\sigma}$ and on the magnetization 
which determines the energy shift of the exchange splitting $\Delta$, see~Eq.~\eqref{exchange}.
%In turn, if density \emph{and} internal energy are given, 
%the temperature and the chemical potential
%can be uniquely determined.
%Both pairs of variables determine the 
%respective other 
%%In turn, temperature and chemical potential
%%determine the density and the internal energy 
%in a unique 
%manner 
The temporal derivatives of the 
energy density $u^{\sigma}_e$ and
%particle density $n^{\sigma}$ as well as of 
the particle density $n^{\sigma}$
%energy density $u^{\sigma}_e$ 
include partial derivatives, e.g.
%, for instance
\begin{align}
\label{eq du}
	\frac{d u^{\sigma}_e}{d t} = c^{\sigma}_T \frac{\partial T^{\sigma}_e}{\partial t}  
		+ c^{\sigma}_{\mu}\, \frac{\partial \mu^{\sigma}}{\partial t}
		+ c^{\sigma}_{m}\, \frac{\partial m}{\partial t}\enspace,
\end{align}
defining the capacity equivalents $c^{\sigma}_x \equiv \frac{\partial u^{\sigma}_e}{\partial x}$.
%according to their variable.
Analogously, partial derivatives of the particle density are defined
as $p^{\sigma}_x \equiv \frac{\partial n^{\sigma}}{\partial x}$.
This allows us to mathematically separate the variables 
$T^{\sigma}_e$, $\mu^{\sigma}$ and $m$. 

%Eqs.~\eqref{eq:N} and \eqref{eq:u} hold separately for 
%majority and minority electrons, defining $n^{\uparrow}$, 
%$n^{\downarrow}$, $u_e^{\uparrow}$ and  
%$u_e^{\downarrow}$.
%The mutually coupled time-dependent variables 
%$T_e^\uparrow, T_e^\downarrow, \mu^\uparrow, \mu^\downarrow, \Delta$
%determine thus the dynamics of energy- and particle exchange 
%of majority and minority electrons,
%and through Eq.~\eqref{exchange} directly the magnetization dynamics. 
%The cooling of the electronic systems plays an important role as
%well, thus the evolution of the lattice temperature $T_{\ell}$ 
%has to be traced simultaneously.

To demonstrate the power of the $\mu$T model and to separate the
time-dependent effects from transport effects,  
we restrict ourselves here to the temporal dependence of the decisive 
variables, %($K=0$)
which is capable to predict and explain
important characteristics of the magnetization dynamics of thin 
ferromagnetic films.
The temporal 
evolution of %the 
%decisive variables 
$T_e^\uparrow, T_e^\downarrow, T_\ell, \mu^\uparrow, \mu^\downarrow$ and $m$
%$T_e^\uparrow, T_e^\downarrow, T_{\ell}, \mu^\uparrow, 
%\mu^\downarrow, \Delta$
is expressed with an equation of 
type \eqref{XTM}:

\begin{align}
\begin{split}
\label{3TM}
\displaystyle 
	\begin{pmatrix}
		c_T^\uparrow & 0 & 0 & c_\mu^\uparrow & 0 & c_m^\uparrow \\
		0 & c_T^\downarrow & 0 & 0 & c_\mu^\downarrow & c_m^\downarrow \\
		0 & 0 & c_\ell & 0 & 0 & 0 \\
		p_T^\uparrow & 0 & 0 & p_\mu^\uparrow & 0 & p_m^\uparrow \\
		0 & p_T^\downarrow & 0 & 0 & p_\mu^\downarrow & p_m^\downarrow \\
		\minus p_T^\uparrow & p_T^\downarrow & 0 & \minus p_\mu^\uparrow & p_\mu^\downarrow & n^\uparrow+n^\downarrow \\
\end{pmatrix}
\frac{d}{dt} 
\begin{pmatrix} T_e^\uparrow \\ T_e^\downarrow \\ T_\ell \\ \mu^\uparrow \\ \mu^\downarrow \\ m \end{pmatrix}
= ~~~~\\
\begin{pmatrix}
		\!\minus\gamma\minus g^\uparrow\! & \gamma & \!g^\uparrow\! & 0 & 0 & 0 \\
		\gamma & \!\minus\gamma \minus g^\downarrow\! & \!g^\downarrow\! & 0 & 0 & 0 \\
		g^\uparrow & g^\downarrow &  \minus g^\uparrow \minus g^\downarrow \!\!\! & 0 & 0 & 0 \\
		0 & 0 & 0 & \minus\nu & \nu & 0 \\
		0 & 0 & 0 & \nu & \!\minus\nu & 0 \\
		0 & 0 & 0 & 0 & 0 & 0 \\
\end{pmatrix}
\!\!\!
\begin{pmatrix} T_e^\uparrow \\ T_e^\downarrow \\ T_\ell \\ \mu^\uparrow \\ \mu^\downarrow \\ m \end{pmatrix}
\!\!+\!\!\begin{pmatrix}
		\! S^\uparrow(t)\! \\
		\!\! S^\downarrow(t)\! \\
		0 \\
		0 \\
		0 \\
		0 \\
\end{pmatrix} 
.
\end{split}
\end{align} 
The first three equations determine the energy of spin-up and spin-down 
electrons as well as of the lattice, respectively. 
Equations four and five trace the densities of both electron systems.
The last equation defines the transient magnetization, Eq.~\eqref{eq:m}. 
In the spirit of the TTM \cite{Anisimov}, 
we introduce an respective equilibration term
for the electron temperatures,
$\pm \gamma\left(T_e^\uparrow-T_e^\downarrow\right)$,
and chemical potentials,  $\pm \nu(\mu^\uparrow-\mu^\downarrow)$.
The laser excitation of each electron system is described by the source term $S^{\sigma}(t)$.
To conserve the total energy with a dynamic exchange splitting,
the correlation energy~\cite{NoltingMag}
%$u_{\rm Corr}\left(T_e^{\uparrow,\downarrow}, \mu^{\uparrow,\downarrow}, m\right)=-U{n^\uparrow n^\downarrow}/\left(n^\uparrow + n^\downarrow\right)$ 
$u_{\rm Corr}(t)=-Un^\uparrow(t) n^\downarrow(t)/n$ 
is taken into account in $c^{\uparrow,\downarrow}_T$, $c^{\uparrow,\downarrow}_\mu$ and $c^{\uparrow,\downarrow}_m$.

\begin{figure}[tb]
	\centering
	\includegraphics[width=\www]
	{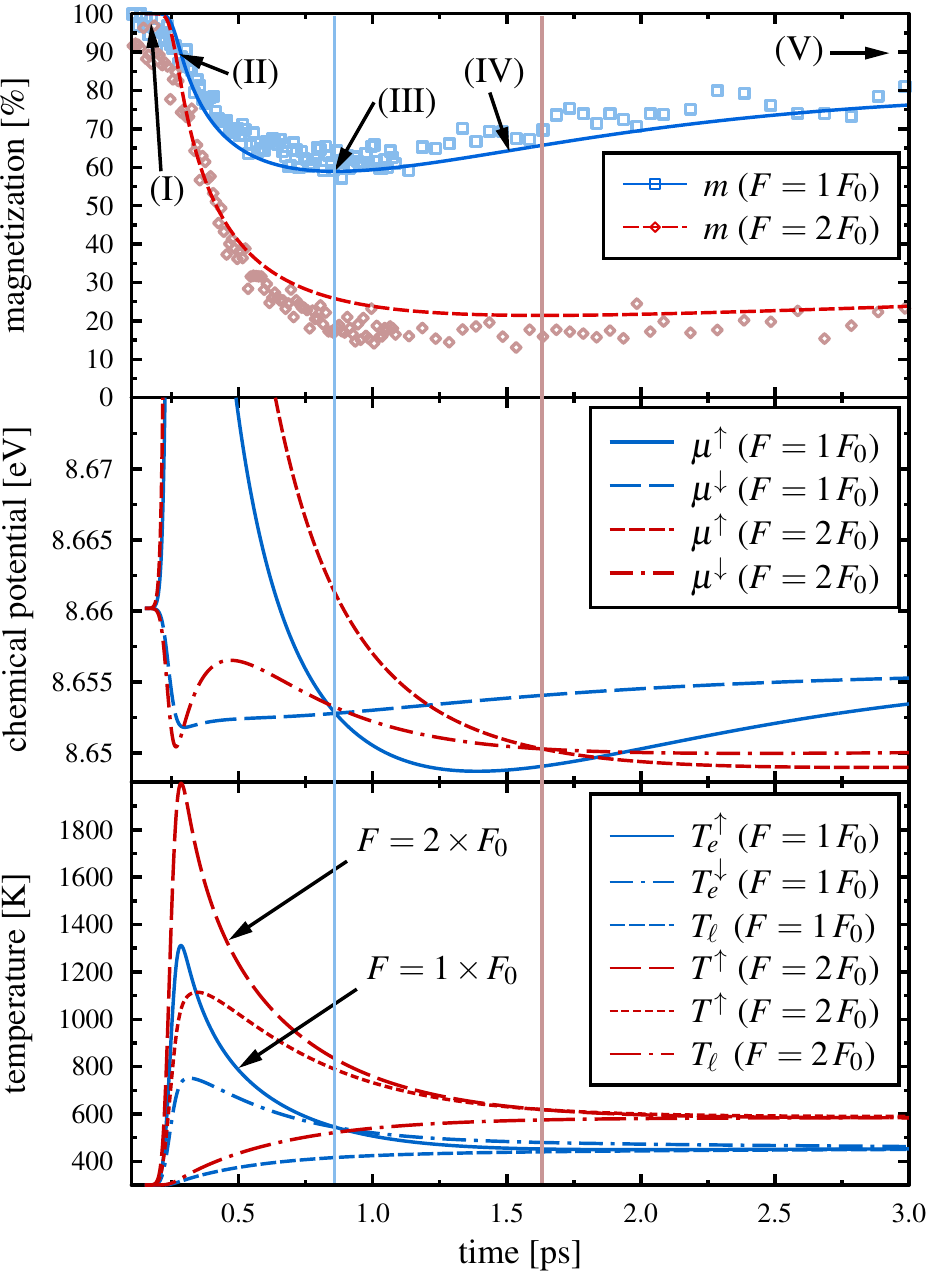}
	\caption{%
	Typical results of the $\mu$T model, 
	transient magnetization (upper panel), 
	chemical potentials (central panel) and 
	temperatures (lower panel). 
	The blue curves correspond to a low fluence 
	$F_0=\unit{2.5}{mJ/cm^2}$, whereas
	the red curves are calculated after excitation with 
	twice of that fluence, $2 \times F_0$.
	In the upper panel, experimental results \cite{Roth2012} 
	are shown for comparison.
	The vertical lines indicate the respective 
	time where the magnetization dynamics suffer a minimum. 
	Characteristic points (I) to (V), as marked for 
	the blue solid demagnetization curve,
	are analyzed in the text.}
	\label{fig:3TM_Mag_TMU}
\end{figure}

We solve the $\mu$TM for nickel, with the density of states 
from  Ref.~\onlinecite{Lin}.
The effective Coulomb interaction $U=\unit{5.04}{eV}$ 
reproduces the experimental \cite{Tyler} equilibrium magnetization curve 
well~\cite{MuellerPRL}.
The lattice heat capacity is taken as $c_\ell = \unit{3.776 \times 10^6}{J/Km^3}$ \cite{CRC}.
For simplicity, we introduce 
the same electron-lattice coupling 
$g^{\sigma} = g/2 = \unit{1 \times 10 ^{18}}{W/Km^3}$
\cite{Mueller2013} for both electron systems.
%For the applied fluences constant material parameters are 
%justified. 
%Their choice is described in the supplementary information.\footnote{Supplementary information is available online.}
The coupling parameters between chemical potentials, $\nu=\unit{5.80 \times 10^{60} }{1/J s m^3}$, and
the inner-electronic temperature coupling, $\gamma=163.8 \times g^{\sigma}$, 
are newly introduced in this work. 
They are determined through a fit of the transient magnetization 
curve obtained by the $\mu$TM to experimental data 
of Ref.~\onlinecite{Roth2012}.
With the same laser parameters as in Ref.~\onlinecite{Roth2012},
and a %\st{fitted} 
reflectivity of $R = 0.44$,
the $\mu$TM reproduces the magnetization curve for 
different fluences.
A comparison between the experiment and the $\mu$T model
is depicted in the upper panel of Fig.~\ref{fig:3TM_Mag_TMU}.

Figure~\ref{fig:3TM_Mag_TMU} 
shows from top to bottom 
the dynamics of the magnetization, of the chemical potentials
$\mu^\uparrow$ and $\mu^\downarrow$ and of the 
temperatures $T_e^\uparrow$, $T_e^\downarrow$ and $T_\ell$.
Two different fluences were applied for the calculations, 
$F_0=\unit{2.5}{mJ/cm^2}$ (blue curves) and $2 \times F_0$
(red curves).
The minima of the magnetization curves
are marked with vertical lines through all three panels of 
Fig.~\ref{fig:3TM_Mag_TMU}.
The chemical potentials (central panel) 
of majority and minority electrons
differ strongly during irradiation, 
equal each other for an instant cross-over and equilibrate on later 
timescales. 
The 
electron temperatures (lower panel) both
grow fast during irradiation, however majority and minority
temperatures differ due to the different heat capacities.
After excitation, both electron temperatures equilibrate with each other 
and later also with the lattice temperature.

Inverting the capacity matrix $C$ in Eq.~\eqref{3TM} leads to a direct formulation 
for the temporal
derivatives of 
$T_e^\uparrow, T_e^\downarrow, T_\ell, \mu^\uparrow, \mu^\downarrow$ 
and $m$. 
%i.e. the variables of $\vec X$, as 
%\begin{align}
%\frac{d \vec X}{dt}=& 
%C^{-1} G \vec X + C^{-1} \vec S
%\label{XTM direct}\enspace.
% \end{align}
In particular,  %the last component %of $\vec X$  
%in \eqref{3TM}
%determines 
%through Eq.~\eqref{exchange}
the change of 
magnetization
is given by
\begin{align}
\label{dmdt}
\frac{dm}{dt}=-\frac{2\nu}{n}\left(\mu^\uparrow-\mu^\downarrow\right)
\enspace,
\end{align}
where the time-dependence occurs only in the difference of the
chemical potentials. Thus, the $\mu$TM directly identifies
the equilibration of chemical potentials of majority and 
minority electrons as 
the driving force of magnetization dynamics,
as proposed in Ref.~\onlinecite{MuellerBY}.

Five characteristic points appear in the magnetization dynamics.
They are indicated in the magnetization curve
for the lower excitation
in 
Fig.~\ref{fig:3TM_Mag_TMU}.
Their origins 
%become now clear 
are explained
with the $\mu$T model
%\red{are}
in the following:

%\begin{enumerate}[noitemsep,nolistsep,itemindent=25pt,leftmargin=0pt,label={(\Roman*)}]
\noindent (I) %\item %1
We analyze the magnetization dynamics directly at the time
when the laser hits the sample. 
Recent ab initio calculations did this as well~\cite{Carva2011,Illg2013},
concluding that the initial change of magnetization, $\left . dm/dt\right|_{t=0}$,
is too small to induce a reasonable demagnetization.
%However, these calculations only treated the very first
%time step and no dynamical exchange splitting was taken into account.~\cite{MuellerPRL}
%
This is in accordance with the $\mu$TM, that predicts even
a vanishing first derivative, $dm/dt$, for the initial time step, 
when the chemical potentials 
are still in equilibrium $\mu^\uparrow=\mu^\downarrow$, see Eq.~\eqref{dmdt}.
%
%The measurement can be taken only after the ice has melted completely.
The feedback effect, induced by a dynamic exchange splitting
only occurs at later times, when the chemical potentials
are driven out of equilibrium.
%
%The laser first has
%to establish a nonequilibrium in chemical potentials before
%the dynamic exchange splitting induces a feedback 
%effect~\cite{MuellerPRL}.
%Since the dynamic exchange splitting changes
The $\mu$TM explicitly accounts for the feedback effect and its influence 
can be illustrated by calculating the second derivative 
%$d^2m/dt^2\propto \nu S(t=0) $
%This can be obtained 
%by taking the derivative 
of Eq.~\eqref{3TM}
during a constant laser excitation
\begin{align*}
\frac{d^2 m}{dt^2}=-
\left( 
\frac{p_T^\uparrow}{c_\mu^\uparrow p_T^\uparrow - c_T^\uparrow p_\mu^\uparrow}
-
\frac{p_T^\downarrow}{c_\mu^\downarrow p_T^\downarrow-c_T^\downarrow p_\mu^\downarrow}
\right)
\frac{\nu S }{n}
\enspace,
\end{align*}
assuming $G$ and $C$ as constant 
over the considered time interval.
Initially, $d^2m/dt^2\propto S(t=0)$
holds for very short times and
the transient magnetization is 
determined mainly by
$m(t)\approx m_0+ \frac 12 \frac{d^2 m}{dt^2} t^2 $.
Thus, even for vanishing
$\left. \frac{dm}{dt}\right|_{t=0}$
the description of demagnetization is possible 
by including a feedback effect,
and ab initio calculations as in Ref.~\cite{Essert2011,Schellekens2013b,Illg2013,Carva2011} 
do not contradict the EY picture.

%
%The last component of the vector $\vec X$ leads to
%\begin{align}
%\displaystyle 
%\!\!\frac{d^2 m}{dt^2}\!&=\!\! 
%\left( 
%\frac{P_T^\downarrow}{C_\mu^\downarrow P_T^\downarrow-C_T^\downarrow P_\mu^\downarrow}
%\!-\!
%\frac{P_T^\uparrow}{C_\mu^\uparrow P_T^\uparrow - C_T^\uparrow P_\mu^\uparrow}
%\right)
%\frac{S \nu}{n}
%% 
%%&\approx 
%%-\unit{6 \times 10^{-29}}{1/m^3} \times  \frac{ S \nu  }{n} 
%\, .
%\label{eq:dm2}
%\end{align}
%
%of the magnetization for short times
%
%However, the initial magnetization dynamics directly due to irradiation
%is determined 
%by the second derivative  
%
%
%Moreover, Eq.~\eqref{eq:dm2} indicates, that the demagnetization effect is directly proportional
%to the laser fluence and also proportional to the coupling $\nu$
%between both chemical potentials. 

%\item %2
\noindent (II) After the excitation, the magnetization decreases 
rapidly, reaching the maximum change at the 
inflection point of the magnetization curve.
%where the change of magnetization is at its 
%maximum.
Eq.~\eqref{dmdt} proposes, 
that also the nonequilibrium of chemical potentials is
at its maximum, which is 
supported
by Fig.~\ref{fig:3TM_Mag_TMU}.

\noindent (III) %\item %3
At the minimum of magnetization 
a transient equilibrium between the electron subsystems is observed.
Here, the
chemical potentials
$\mu^\uparrow=\mu^\downarrow$ (as expected from Eq.~\eqref{dmdt})
and also 
the temperatures $T_e^\uparrow=T_e^\downarrow$
are equilibrated, both confirmed by Fig.~\ref{fig:3TM_Mag_TMU}.
However, the lattice is still not in equilibrium with the electron system.
In this  transient equilibrium state,
the $\mu$TM shows
that the parabola approximation of the minimum,
\begin{align*}
\frac{d^2 m}{dt^2} =
\frac{2 g \nu}{n} \left( 
\frac{p_T^\uparrow}{c_\mu^\uparrow p_T^\uparrow - c_T^\uparrow p_\mu^\uparrow}
-
\frac{p_T^\downarrow}{c_\mu^\downarrow p_T^\downarrow-c_T^\downarrow p_\mu^\downarrow}
 \right) 
\left (T_e-T_\ell\right),
\end{align*}
is mainly determined by 
the temperature difference between the electrons and the lattice.

%\item 
\noindent (IV)
After the transient equilibrium state of the electron subsystems, 
the chemical potentials are driven out of equilibrium
again. This is due to the relaxation with the lattice. 
At the maximum difference between both 
chemical potentials, the second inflection point in the magnetization 
curve occurs.

\noindent (V) %\item 
For larger times, the chemical potentials 
and temperatures of the electrons and the lattice 
equilibrate,
%are in equilibrium, 
see Fig.~\ref{fig:3TM_Mag_TMU},
and the magnetization reaches its equilibrium value 
$m\left(T_e\right)$. 
%\end{enumerate}

\begin{figure}[tb]
	\centering
	\includegraphics[width=\www]
	{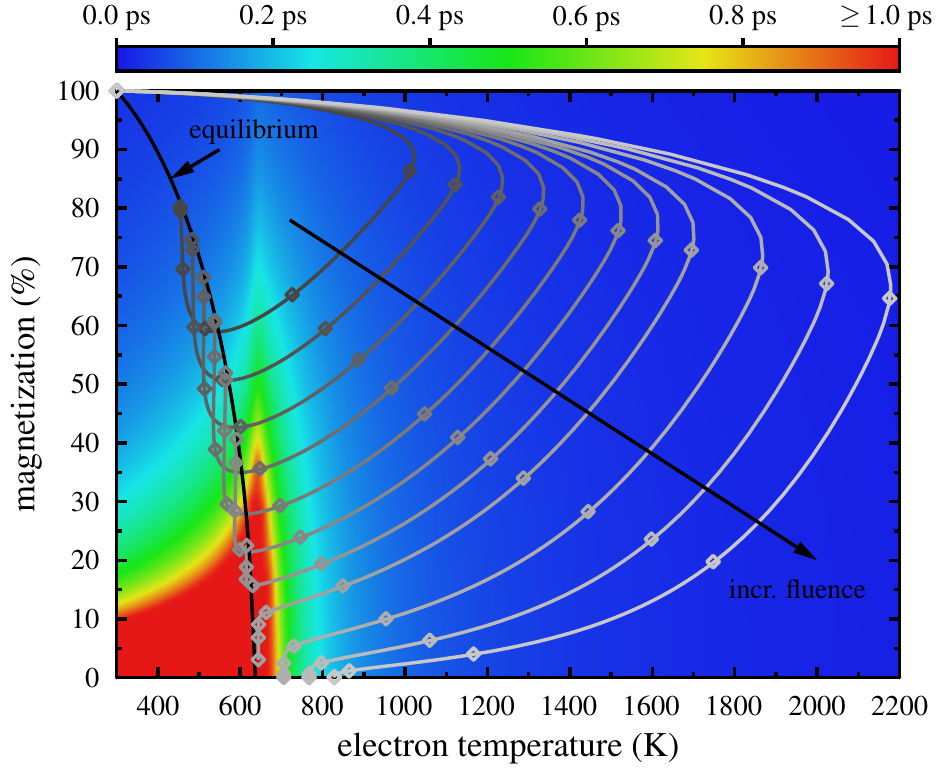}
	\caption{Phase diagram of ultrafast magnetization dynamics.
	The black curve is the equilibrium magnetization cuve $m(T)$.
	The gray curves result from the $\mu$T model for 
	the fluences 
	$F/F_0 = 1.0,1.2,...,2.4, 2.8, 3.2,3.6$
	with $F_0=\unit{2.5}{mJ/cm^2}$.
	The dots mark several times at $t=\unit{0, 0.3, 0.5, 1, 2, 5, 10, 25}{ps}$.		
	The background color labels the relaxation time towards the equilibrium magnetization.
	%Its color code is given on top of the diagram.
	}
	\label{fig:phase}
\end{figure}

%\section{Phasendiagramm}
The strength of the $\mu$TM is the possibility of
analytical predictions about many relevant 
physical processes in ultrafast magnetization dynamics. 
In particular, we observe in Fig.~\ref{fig:3TM_Mag_TMU} a 
so-called \emph{critical slowing down}
of magnetization dynamics 
 \cite{Chubykalo-Fesenko2006,Munzenberg2010} 
for the high laser fluence.
%which shows 
%\red{meaning}
%a much slower than for the lower fluence.
The reason is explained with Fig.~\ref{fig:phase},
which depicts the phase diagram of $m$ and $T_e$.
For long excitations, in the order of nanoseconds, 
we expect that the magnetization 
follows the equilibrium magnetization $m\left(T_e\right)$ which is indicated as a black curve.
However, the ultrashort laser pulse drives the system out of equilibrium
and the magnetization 
becomes a function of %$m\left(
$T^\uparrow_e,T^\downarrow_e,\mu^\uparrow$ and $\mu^\downarrow$.
%\right)$. 
In particular, in these nonequilibrium states, the chemical potentials differ strongly,
which is reflected in the central panel of Fig.~\ref{fig:3TM_Mag_TMU}.
The temperatures $T_e^\uparrow$ and $T_e^\downarrow$
are close to each other and 
are approximated by their mean value $T_e$ for the following discussion.
In Fig.~\ref{fig:phase}, the parametric $\left(m,T_e\right)$ curves 
of magnetization dynamics after different 
laser fluences are indicated by 
gray lines.
All curves start at room temperature, top left of the diagram.
Further dots up to  \unit{25}{ps} show the dynamical behavior on the parametric curves.
The laser drives the system to high electronic temperatures, 
however, due to the nonequilibrium situation, the magnetization is still finite even 
for $T>T_C$.
The first cross-over of the equilibrium magnetization 
curve, observed
for fluences up to $2.2 \times F_0$, corresponds to the \emph{transient
equilibrium state} (III) and coincides with the minimum of the 
respective magnetization curve.
%The 
%equiblibrium magnetization $m(T_e)$
%coincides with the minima of the
%laser-excited magnetization curves,
%indicating the
%transient equilibrium state.
%

Importantly, the time to reach the final state on the equilibrium magnetization 
curve $m(T_e)$ differs for different fluences. 
For each pair $\left(m,T_e\right)$ both 
chemical potentials $\mu^\uparrow,\mu^\downarrow$ can be 
determined by 
simultaneously
solving Eq.~\eqref{eq:m}
and the equation of particle conservation, $n={\rm const}$. 
We can estimate the time $\tau_{\rm eq}$ to reach the equilibrium 
magnetization (black curve) for each point of $m$ and $T_e$ in the phase diagram
by a relaxation time approximation
of Eq.~\eqref{dmdt},

\begin{align}
\label{eq:relax}
\frac{m\left(T_e,\mu^\uparrow,\mu^\downarrow\right)-m\left(T_e\right)}{\tau_{\rm eq}}
=-\frac{2\nu}{n}\left(\mu^\uparrow-\mu^\downarrow\right)
\enspace.
\end{align}
The relaxation time to equilibrium, $\tau_{\rm eq}$,  is depicted in the background color code of Fig.~\ref{fig:phase}.
Under strong
nonequilibrium conditions, especially at high temperatures,
this relaxation occurs very fast: The large
difference in 
chemical potentials rapidly drives the magnetization to its equilibrium value.  
However, around the Curie temperature at \unit{631}{K} \cite{CRC}, 
the chemical potentials are nearly equal and the equilibration
time according to Eq.~\eqref{eq:relax} reaches rather high values up to nanoseconds, thus,
the magnetization dynamics is extremely decelerated.
%\red{\st{In Fig.2 certain time steps are marked to illustrate the dynamical behavior.}}
The fluences $F/F_0\approx 2.0-2.4$ drive the system
into this critical region, appearing red in Fig.~\ref{fig:phase}.
%,  
%and even after \unit{10}{ps} 
%the magnetization has not reached its equilibrium value.
%\red{continuing} changes \st{in the magnetization} are observed. 
%
For low and very high laser fluences this region is circumvented.
%With this analysis, we conclude, that the slowing down is mainly governed 
%by the difference in chemical potentials. 
%
%The M3TM also reproduces this experimentally observed slowing down. 
Thus, the $\mu$TM directly illustrates the %transition towards
origin of
a critical slowing down  
and explains why experiments 
show a maximum in demagnetization time~\cite{Koopmans2010},
by
%\red{as well as of non-monotonous demagnetization times %\cite{Koopmans2010}}
utilizing basic  
thermodynamical concepts.
%\pink{utilizing:. Unten nur ''by''}

%\section{Conclusion}
In conclusion, we derived the $\mu$T model for 
itinerant ferromagnets.
The description traces
the dynamics of spin-resolved 
electron
temperatures \emph{and}  chemical potentials simultaneously
and combined with the coupling to the lattice.
The demagnetization process can be described 
based on a few fundamental physical concepts, like dynamic exchange splitting 
and the 
relaxation towards thermodynamic equilibrium.
%equilibration of chemical potentials 
%and temperatures.
%We conclude, that ab initio calculations
%utilizing the same chemical potentials for majority
%and minority electrons with a fixed band structure
%do not contradict the Elliott-Yafet type picture.
%of majority and minortiy electrons
%
Our method identifies the
minimum of the magnetization 
as a transient equilibrium state of the electron systems.
%\red{\st{We have demonstrated the possibility to analyze ultrafast demagnetization
%due to the nonequilibrium of temperatures and chemical potentials. }}
%We conclude, that ab initio calculations,
%as in Ref.~\cite{Carva2011,Illg2013}, do not contradict the 
%Elliott-Yafet type picture.
%With our manuscript we intend to encourage further ab initio
%calculations which explicitly take the nonequilibrium of chemical potentials
%and the feedback effect into account. 
%\st{The main statement of this Letter}
%the $\mu$T model 
We
explain 
%the dramatic deceleration of the magnetization dynamics
the experimentally observed slowing down of the magnetization dynamics
%in ultrafast magnetization by 
by %the entering of 
a critical region in the magnetic phase diagram, Fig.~\ref{fig:phase}.
For certain fluences, the system is driven into this region 
and the time to reach the equilibrium magnetization increases considerably.
%\red{The extension of the $\mu$TM to account for superdiffusive transport
%effects is subject of current research.}

%\pink{Bin gerade nicht sicher, ob mit dem Satz und dem im Abstract 
%ich den Transport jetzt nicht {\em noch} mehr betont habe. Aber da die 
%Einleitung sich daran aufzieht, muessen wir da schon irgendwie drauf 
%achten, bzw. uns entscheiden, ob wir das mit verkaufen wollen, oder eben
%nicht.}

%\section{Acknowledgments}
Financial  support  of
the  Deutsche
Forschungsgemeinschaft  through the Heisenberg project RE 1141/15
``Ultrafast Dynamics of Laser-excited Solids''
is gratefully acknowledged.
%\pink{Dank an MAe und HCS for helpful discussions?}
%\vspace*{-3em}

%\bibliographystyle{prl_etal}
%\bibliographystyle{apsrev}
%\bibliography{library} 

%Merlin.mbs v4.21 2009-07-09.
%

\end{document}